# Compensatory Mechanisms in Non-principal Multimedia Learning: The Interplay of Local and Global Information Processing

Mohammadhosein Ostadi Varnosfaderani [1], Masoumeh Golmohamadian [2], Alireza Bosaghzadeh [3], S. Hamid Amiri [3] & Reza Ebrahimpour [4*]

[1] Department of Electrical Engineering, Sharif University of Technology, Tehran, Iran
[2] School of Cognitive Sciences (SCS), Institute for Research in Fundamental Science (IPM), Tehran, Iran
[3] Department of Computer Engineering, Shahid Rajaee Teacher Training University, Tehran, Iran
[4] Center for Cognitive Science, Institute for Convergence Science and Technology (ICST), Sharif University of Technology, Tehran P.O. Box:11155-1639, Iran
*Correspondence and requests for materials should be addressed to R.E.(email: ebrahimpour@sharif.edu)*

## Abstract

Educational multimedia has become increasingly important in modern learning environments because of its cost-effectiveness and ability to overcome the temporal and spatial limitations of traditional methods. However, the complex cognitive processes involved in multimedia learning pose challenges in understanding its neural mechanisms. This study employs network neuroscience to investigate how multimedia design principles influence the underlying neural mechanisms by examining interactions among various brain regions. Two distinct multimedia programs were constructed using identical auditory content but differing visual designs: one adhered to five guidelines for optimizing multimedia instruction, referred to as principal multimedia, while the other intentionally violated these guidelines, referred to as non-principal multimedia. Cortical functional brain networks were then extracted from EEG data to evaluate local and global information processing across the two conditions. Network measurements revealed that principal networks exhibited more efficient local information processing, whereas non-principal networks demonstrated enhanced global information processing and hub formation. Network modularity analysis also indicated two distinct modular organizations, with modules in non-principal networks displaying higher integration and lower segregation than those in principal networks, aligning with initial findings. These observations suggest that the brain may employ compensatory mechanisms to enhance learning and manage cognitive load despite less effective instructional designs.

**Keywords:** Educational multimedia, Cognitive load, EEG, Functional brain networks, Global and local information processing

# Introduction

When we want to learn about a topic, many people turn to educational multimedia, which frequently offers an efficient and convenient solution, especially when we consider time limitations and budget constraints. Teachers also typically incorporate educational multimedia into their lessons, expecting students to learn from the content presented. However, the learner's experience can vary greatly depending on the design and content of the multimedia [1–3]. In some cases, learners easily follow the multimedia program and use it as a foundation for further learning, engaging with the content and achieving better learning outcomes. Conversely, learners may struggle to engage with poorly designed videos and feel anxious and exhausted, leading to an inability to apply the content later. This raises important questions: what is the neural mechanism responsible for governing our internal evaluations and performance during and after watching educational multimedia? How do different brain regions interact to facilitate more effective learning while watching a video?

Learning from educational multimedia is a complex cognitive task that occurs in a multidimensional space, resulting in varying levels of cognitive workload in the brain. How the human brain processes and stores information has been discussed in cognitive load theory (CLT) [4]. In modern education, CLT provides valuable insights for designing efficient educational multimedia that optimizes the cognitive load and improves the learning process [3,5,6]. In 2002, Richard Mayer introduced twelve principles to improve multimedia design and reduce the cognitive load for learners [2]. Five principles specifically target extraneous processing, which involves cognitive effort on nonessential elements. These include coherence (removing irrelevant content), signaling (highlighting key information), redundancy (excluding on-screen text when graphics and narration are used), spatial contiguity (placing related words and images close together), and temporal contiguity (presenting related words and images simultaneously). Adhering to these 5 principles helps create more effective multimedia learning experiences with optimized cognitive load [7–9].

Like other cognitive tasks, assessing the cognitive workload during the learning process requires accurate and reliable measurement techniques. Researchers commonly employ two main approaches: subjective and objective. Subjective methods involve questionnaires and interviews to evaluate cognitive load based on the subject's perception of task difficulty[1,10], such as the NASA-TLX questionnaire [11]. Objective methods, on the other hand, utilize physiological and behavioral measures, including electroencephalography (EEG) [12,13], eye-tracking [14,15], and functional magnetic resonance imaging (fMRI) [16]. By employing both subjective and objective measures, we can gain a more comprehensive view of the complex interplay between neural processes and their manifestations in human experience and behavior [17].

Among all these methods, EEG, either alone or in conjunction with other approaches, plays a key role in analyzing brain function because of its real-time monitoring ability, long-term recording potential, and high temporal resolution [18]. Researchers have employed various statistical analyses [19–21] and machine learning techniques [8,22] on EEG data to evaluate different aspects of cognitive load in the learning process, including memory and attention. Furthermore, substantial research has been conducted on frequency band analysis, with a particular focus on the theta and alpha bands [23,24]. Changes in theta frequency band activity are associated with working memory performance, particularly in frontal and central brain regions [19,25]. Moreover, theta band power increases during successful memory encoding [26], and theta band activity is positively correlated with memory retrieval [27]. However, alpha-band oscillations are correlated with attention and the control of access to stored information, particularly in the parietal and occipital lobes [28]. EEG alpha power has also been shown to be associated with creative ideation[29]. However, a crucial aspect that remains unclear is how different brain regions interact with each other across various frequency bands, leading to varying levels of cognitive workload.

Recent research suggests that our understanding of the principles and mechanisms underlying complex brain function can be improved by integrating data from multiple brain regions and examining their functional interactions. Rather than focusing solely on localized changes in brain structure and function [30,31], network neuroscience takes an integrative approach. It models the brain as a complex network, with brain regions as nodes and their connections as edges, allowing the application of graph theory and network analysis to study the brain's functional mechanism at the whole-brain scale [32–34]. In this way, several network measurements have been employed to characterize one or several aspects of global and local functional brain connectivity. Three common types of measurements are as follows: a) Segregation measurements reveal that the brain network exhibits a modular organization, with densely connected clusters of brain regions responsible for specialized information processing [35]. b) Integration measurements: The modular architecture of the brain network resembles a highly efficient "small-world" structure, striking a balance between local clustering and global integration. The integration measures evaluate the global information processing in the brain [35,36]. c) Centrality measurements: These identify "hub" regions that play crucial roles in integrating information across distributed brain systems [37].

Adopting the network neuroscience approach, which models the brain as a complex network of interconnected regions, can provide valuable insights into the neural mechanisms underlying multimedia learning and its cognitive workload. Recent studies have revealed that different factors can affect the pattern of the brain's connectivity while watching a video, for example, receptive versus interactive video screens [38] and shape color [39].

This study aims to investigate how Mayer's principles mentioned above can affect functional brain networks to facilitate and improve the learning process and optimize cognitive workload. To achieve this goal, we constructed two educational multimedia programs with the same auditory content; however, in one, the visual design adhered to Mayer's principles, whereas in the other, it violated these principles. We employed both subjective and objective methods to assess the impact of these multimedia designs on learning outcomes and cognitive load. The subjective measures included a recall test and the NASA-TLX questionnaire, whereas the objective measures involved analyzing brain networks obtained from the EEG data. We examined local and global information processing in the brain across two conditions via relevant functional brain network measurements, network modularity and behavioral outcomes separately and explored the relationships between them.

## Materials and methods

### Participants
The dataset included 39 healthy adult volunteers who were local university students, aged between 20 and 29 years, with a mean age of 22.8 years and a standard deviation of ±2.5 years. Some participants were excluded because of incomplete recordings (n = 2), noisy data (n = 2), or low post-test scores (n = 1). For additional details, refer to earlier research on this dataset [7,8,40]. Two educational multimedia learning videos were designed to investigate the impact of Mayer's multimedia learning principles on the neural mechanism of the brain. One of these videos followed Mayer's principles (principal multimedia or P), whereas the other one intentionally avoided these principles (non-principal videos or NP). Crucially, the auditory content of the two videos was the same, but the visual content was different. Sixteen participants learned materials from the principal video and the remaining 18 participants learned from the non-principal videos.

All the participants were right-handed, which helped avoid hemispheric lateralization. Additionally, they had normal hearing and normal or corrected-to-normal vision, with no history of head injury. Their primary language was Persian, and they were also proficient in English. Before they participated in the study, they completed a standard pre-task listening test and a similar test to familiarize themselves with the main task. Informed written consent was obtained from all participants. The study followed approved protocols from the Iran University of Medical Sciences (IR.IUMS.REC.1397.951) and adhered to the guidelines and regulations outlined by the Declaration of Helsinki.

### Computer-based educational multimedia

In this study, one listening chapter, Lesson 11, from Open Forum 3 [41], was chosen as the core material for developing two educational multimedia programs. This selection is accessible online through Oxford University Press (https://elt.oup.com/student/openforum/3?cc=ir&selLanguage=en). The listening chapter served as the foundation for creating two distinct versions of multimedia: principal and non-principal. A motion graphics expert utilized Adobe After Effects CC 2017 (version 14.2.1.34) to create the multimedia content. The video duration for lesson 11 is 342 seconds. The two multimedia programs used in this experiment are available on GitHub (https://github.com/K-Hun/multimedia-learning-hci).

### Experimental Task

The participants were seated on an adjustable chair, 57 cm away from a 17-inch monitor with a refresh rate of 60 Hz and in a partially sound-attenuated and dim-light room. To reduce head movements and ensure consistent data collection, the subjects put their head on a chin rest. In front of the subject, there were two loudspeakers, one on the right and the other on the left. Each participant was presented thoroughly about the experiment and the procedure, which involved four distinct sessions: (1) a resting-state session where participants looked at a black-filled circle, (2) a multimedia learning session without interaction, (3) a recall test conducted via a mouse interface, and (4) completion of the NASA-TLX questionnaire via a paper-based version. Electroencephalography (EEG) signals were collected during the first three steps. Since this study focused on analyzing functional brain networks during the learning process, we analyzed only session 2 (the learning task). Figure 1 illustrates the experimental design.

At the beginning of the experiment the EEG cap was set up, and the recording began. After a few seconds, when a timer ended, the multimedia program automatically played. The participants were aware that they focused on the concepts presented. There was no interaction during the video. After the multimedia program, a recall test started automatically. The participants could answer questions via a mouse interface, leaving any questions unanswered. They could move between questions and terminate the test before the timer ended. Following these steps, the EEG stopped, and a paper-based NASA-TLX was given [42,43] to ensure that the two versions of the multimedia presentations were different in their levels of cognitive load. As previously mentioned, the subjects were randomly divided into two groups for this study. The first group is exposed to Lesson 11 NP, whereas the second group views Lesson 11 P.

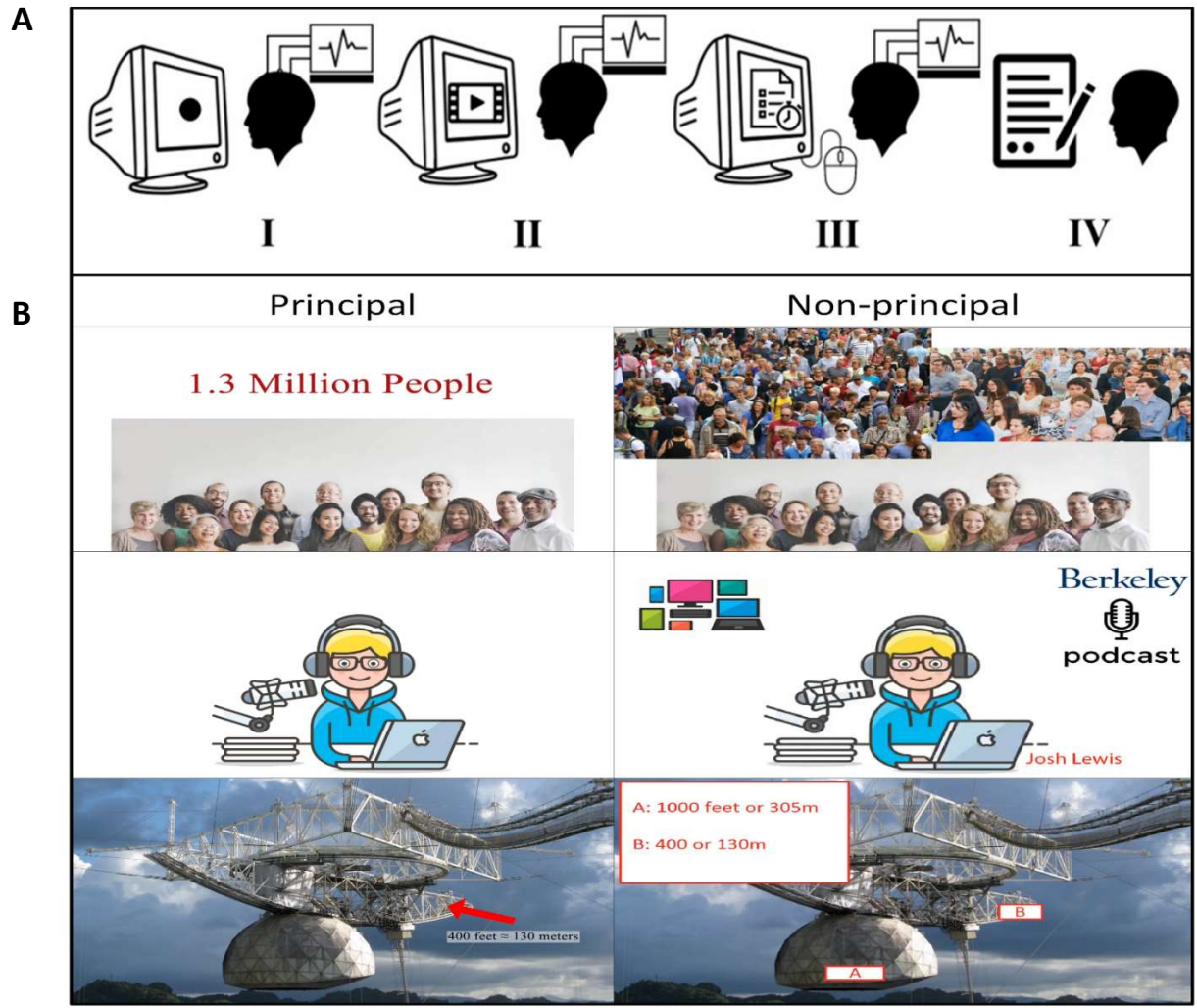

**Figure 1. Experimental procedure and multimedia design**
(A) The experiment was conducted in a systematic sequence (from left to right): participants began by observing a black-filled circle to gather baseline data. The participants were subsequently presented with multimedia content without any interaction. Following this, the participants engaged in a recall test utilizing a mouse interface. Finally, they completed the NASA-TLX questionnaire in paper format. EEG signals were collected during the first, second, and third phases of the experiment. (B) Three example frames of the principal design and non-principal design of multimedia: both use the same auditory content.

### EEG Data Collection and Preprocessing

For EEG data collection, a portable 32-channel eWave amplifier [44–46] paired with eProbe v6.7.3.0 software was used. The data were recorded via 29 passive wet electrodes placed on the scalp according to the 10–20 system[7,8,40]. Additionally, we used bilateral mastoids (M1 on the left and M2 on the right) as reference points for the EEG signals. The electrode topography was organized as follows (Figure 2): the prefrontal cortex (PFC) includes Fp1 and Fp2; the midline prefrontal cortex (mPFC) is represented by Fpz; the ventrolateral prefrontal cortex (VLPFC) comprises F7 and F8; the dorsolateral prefrontal cortex (DLPFC) consists of F3 and F4; the frontal cortex (FC) encompasses FC5, FC1, FC2, and FC6; the midline frontal cortex (mFC) includes Fz and Cz; the temporal cortex (TC) contains T7, T8, P7, and P8; the parietal cortex

(PC) is made up of C3, C4, CP5, CP1, CP2, CP6, P3, and P4; the midline parietal cortex (mPC) is denoted by Pz; the occipital cortex (OC) involves O1 and O2; and the midline occipital cortex (mOC) is indicated by POz. The system records data with 24-bit resolution at a rate of 1,000 samples per second. Visual triggers on the monitor were also employed to ensure synchronization. Electrode impedances were kept below 5 $K\Omega$ in all recordings and electrode sites.

EEG data analysis and preprocessing were carried out via EEGLAB[47] Toolbox version 2020.0 in the MATLAB environment. Before performing EEG data preprocessing, we interpolated the missed channels via the spherical spline interpolation method[48,49]. By incorporating this interpolation step early in the preprocessing pipeline, we maintain the integrity of the EEG data, which is essential for obtaining accurate results[50].

To begin the preprocessing of EEG signals, we applied the basic FIR band-pass filter within the 0.5-48 Hz range to eliminate DC and high-frequency noise. However, mastoid referencing introduces external experimental artifacts in EEG signals because of unstable connections with the mastoids. To mitigate this effect, we employed the re-referencing component of the PREP pipeline algorithm [51] to estimate the true reference. Additionally, we utilized the artifact subspace reconstruction (ASR) algorithm [52] to correct corrupted segments of the EEG data, including the removal of high-amplitude components such as eye blinks, muscle movements, and sensor motion [53]. We perform ASR using the *Clean_Rawdata* plug-in with default settings. In the final preprocessing step, we applied independent component analysis (ICA) via the fastICA algorithm to remove the remaining artifacts (specifically, eye movements) from the data.

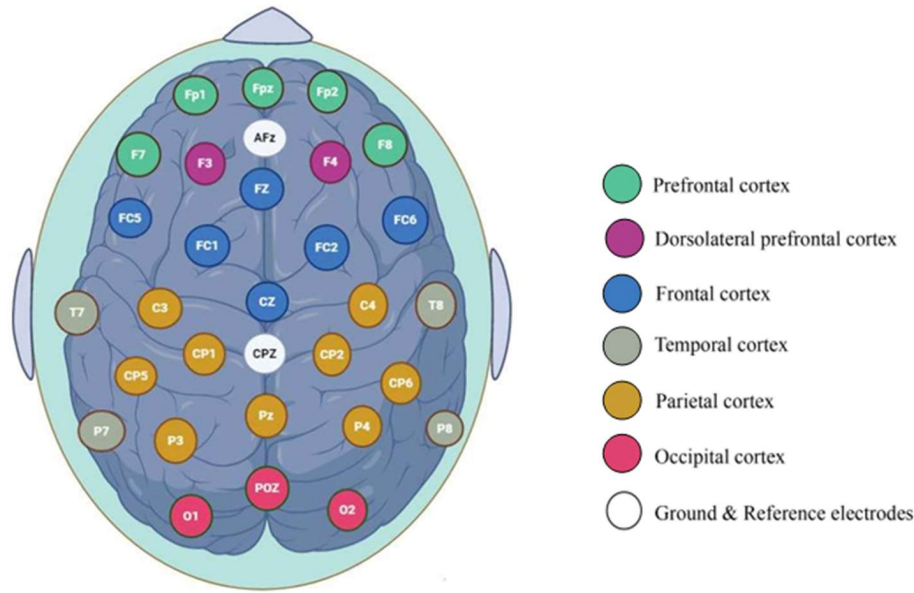

**Figure 2. Electrode placement using the extended international 10–20 system**
The extended international 10- 20 system (10% system) was employed to place 32 electrodes on the scalp, with the electrodes distributed across various regions of the cortex. The electrodes were categorized as follows: prefrontal (Fp1, Fp2), medial prefrontal (mPFC) at Fpz, ventrolateral prefrontal (F7, F8), dorsolateral prefrontal (F3, F4), frontal (FC5, FC1, FC2, FC6), midfrontal (Fz, Cz), temporal (T7, T8, P7, P8), parietal (C3, C4, CP5, CP1, CP2, CP6, P3, P4), midparietal (Pz), occipital (O1, O2), and midoccipital (POz). The reference for the system was CPz, with grounding at AFz.

**Surface Laplacian Transformation**

Source imaging algorithms, such as weighted minimum-norm estimation (wMNE), exact low-resolution electromagnetic tomography (eLORETA), and beamforming, aim to mitigate the effects of volume conduction. However, these methods have limitations. Previous studies have indicated that using a smaller number of electrodes than the minimum requirement (64 electrodes) can lead to inaccurate source reconstruction[54]. Moreover, the inverse problem lacks a singular solution[55–57] because of the demands of source reconstruction algorithms, which necessitate precise inverse and forward models, selection of anatomical templates and head volume conductor models (including tissue conductivities), and initial assumptions about the sources. To address this challenge, the surface laplacian (SL) provides an alternative approach for estimating current-source density (CSD) that offers several advantages. SL does not require assumptions about the sources, which makes it reference-free. Additionally, SL can produce reliable results even when a low-density electrode setup with fewer than 64 electrodes is used [48,58]. In our study, we utilized the surface laplacian (SL) on the corrected EEG signals to estimate the current-source density (CSD). By adopting this approach, we aimed to improve topographical localization and minimize volume conduction effects. A similar approach was employed in other studies [39,59,60]. Finally, for a more detailed examination of EEG signals across diverse spectral ranges, we considered corrected EEG signals in five frequency bands: delta (0.5–4 Hz), theta (4–8 Hz), alpha (8–13 Hz), beta (13–30 Hz), and gamma (30–48 Hz).

**Brain network construction**

In our study, we employed the phase slope index ($PSI$) to quantify the magnitude and direction of information flow between different brain regions[61]. The implementation of PSI is publicly available at http://doc.ml.tu-berlin.de/causality/. $PSI$ was selected for its insensitivity to volume conduction artifacts and its ability to identify non-zero phase delays, facilitating the accurate estimation of effective connectivity networks (ECNs) at the sensor level. In this method, the estimated constant time delay ($\tau$) is used to study the interaction between different brain regions. More specifically, ($\tau$) is a linear phase shift in a frequency domain. Importantly, the phase slope of the cross-spectra, denoted as, $\Phi(f)$ is also a function of frequency, which is expressed as:

$$\Phi(f) = 2\pi f \tau$$

The phase slope sign, positive or negative, indicates the direction of interaction, such as between electrodes $i$ and $j$. If the slope is positive, it suggests a causal relationship from $y_i$ to $y_j$ (with $i$ as the origin and $j$ as the destination). Conversely, a negative slope implies that the interaction flows from $y_j$ to $y_i$. In general, $PSI$ is calculated as follows:

$$\widetilde{\Psi}_{i,j} = \Im\left(\sum_{f \in F} Coh_{ij}^*(f) Coh_{ij}(f + \delta f)\right)$$

where $F$ is the set of all frequencies; $\Im(.)$ is the imaginary part of coherency; $Coh_{ij}(f) = S_{ij}(f)/\sqrt{S_{ii}(f)S_{jj}(f)}$ is the complex coherency; $\delta f$ is the frequency resolution and $\delta f = F_s/n_{FFT} = 0.5\ HZ$; and $S$ is the cross-spectral matrix. Finally, the $PSI$ values are normalized, which is calculated via the Jackknife method [61].

$$\Psi = \widetilde{\Psi}/std(\widetilde{\Psi})$$

Importantly, the weighted average of the slope, $\widetilde{\Psi}$, becomes zero if the imaginary part of the coherency approaches zero. This makes it insensitive to mixtures of non-interacting sources, as it is unaffected by zero phase differences, making it a robust measure of interactions between brain regions [62]. In the next step, the normalized $PSI$ are sorted in a $29 \times 29$ skew-symmetric matrix, including 406 possible pairwise associations ($(N^2 - N)/2$, where $N = 29$). Then, a statistical threshold ($|\Psi| > 2$) corresponding to a 95% confidence interval of the $PSI$ distribution at $p < 0.05$ (two-tailed) was applied to the obtained matrix [61]. This threshold was used to prepare the data for graph theoretical analysis where values falling below the threshold were reassigned to zero, whereas those surpassing the threshold maintained their original values. To extract effective connectivity networks and apply graph theoretical analysis, we relied on the obtained directed and weighted adjacency matrix. In this context, electrodes serve as nodes within the graph or network, and the edges represent connections between different brain regions, with each edge corresponding to an entry in the adjacency matrix.

### Measures of brain networks

The topological properties of identified effective connectivity networks can be analyzed via fundamental graph theoretical measures such as the node degree and directionality index [33]. Additionally, key metrics related to integration, segregation, network sparsity and centrality play crucial roles in understanding the network's structure. We computed all these metrics from the directed and weighted adjacency matrices via the Brain Connectivity Toolbox (BCT)[32].

### Node degree and directionality index

The degree of a node refers to the number of edges connected to that node. Since an adjacency matrix is generated from $PSI$ values, it reveals the direction of information flow. To compute a node's degree, we divide it into two components: The in-degree ($k_i^{in}$), which represents the strength of the incoming flow, and the out-degree ($k_i^{out}$), which represents the strength of the outgoing flow and are computed as follows:

$$\begin{cases} k_i^{in} = \sum_{j \in N} A_{ji} \\ k_i^{out} = \sum_{j \in N} A_{ij} \end{cases}$$

Nodes with high out-degree values are regions that can exert influence over others. Conversely, nodes with high in-degree values indicate areas influenced by other regions. In this way, the total degree ($TD$) characterizes the hubness or centrality of a node in the network. Mathematically, $TD$ is expressed as

$$TD = \sum k_i^{in} + \sum k_i^{out}$$

To determine the overall direction of the information flow in each node, we can use the directionality index ($DI$). This measurement is expressed as

$$DI = \sum k_i^{out} - \sum k_i^{in}$$

A positive value of $DI$ indicates that the electrode behaves as a source or sender of information, actively transmitting data. Conversely, a negative value of $DI$ denotes the electrode's role as a sink or receiver, passively accepting incoming information.

**Measures of functional segregation**

The brain exhibits functional segregation, allowing specialized processing to occur in densely interconnected groups of brain regions. Measures of segregation primarily evaluate the presence of these groups, which are commonly known as clusters or modules, within the network. The clustering coefficient ($CC$) of a node represents how close its neighbors tend to cluster together [63]. To compute the clustering coefficient, we average all local clustering coefficients, resulting in a value between 0 and 1. Mathematically, it is computed as follows:

$$CC = \frac{1}{N}\sum_{i \in N} C_i = \frac{1}{N}\sum_{i \in N} \frac{t_i}{(k_i^{out} + k_i^{in})(k_i^{out} + k_i^{in} - 1) - 2\sum_{j \in N} A_{ij}A_{ij}}$$

where $N$ is the number of nodes; $C_i$ is the local clustering coefficient; $t_i$ is the number of triangles that exist around each node; $k_i^{in}$ and $k_i^{out}$ are the in-degree and out-degree of a node, respectively; and $A_{ij}$, $A_{ji}$ where is the entry of the adjacency matrix.

Local efficiency ($LE$) quantifies how information is transmitted within local clusters. It reflects the ability of neighboring nodes to communicate efficiently when a specific node is removed. High local efficiency values facilitate parallel processing, allowing effective integration of information.

$$LE = \frac{1}{2N}\sum_{i \in N} \frac{\sum_{j,h \in N, j \neq i}(A_{ij} + A_{ij})(A_{ih} + A_{hi})\left[\left(d_{jh}(N_i)\right)^{-1} + \left(d_{hj}(N_i)\right)^{-1}\right]}{(k_i^{out} + k_i^{in})(k_i^{out} + k_i^{in} - 1) - 2\sum_{j \in N} A_{ij}A_{ij}}$$

where $d_{hj}(N_i)$ is the length of the shortest path between $j$ and $h$ that contains only neighbors of $i$.

**Measure of functional integration**

Functional integration in the brain denotes the ability to rapidly merge specialized information from diverse brain regions. Measures of integration define this concept by assessing how easily brain regions communicate. These measures typically rely on the concept of a path, which consists of sequences of separate nodes and links [33]. The global efficiency ($GE$) is a fundamental indicator of a network's integration and is computed by the average inverse shortest path length [64]. The shortest path length ($PL$) is the lowest number of edges that exist between any given pair of nodes.

$$GE = \frac{1}{N}\sum_{i \in N}(\sum_{j \in N, j \neq i}(PL)^{-1}/N - 1)$$

High $GE$ values indicate efficient communication and fewer processing steps between network nodes, whereas low $GE$ values indicate the opposite.

**Measures of centrality and network sparsity**

Important brain regions, known as hubs, communicate with many other regions, enabling functional integration and contributing significantly to network resilience against insults. Different measures of centrality assess the significance of individual nodes based on these principles [33]. Betweenness centrality ($BC$) represents the proportion of all the shortest paths in a network that path to a specific node. Nodes with high $BC$ values serve as hubs, participating in numerous "shortest" paths. Mathematically, $BC$ is computed for each node in the network by

$$BC_i = \frac{1}{(N-1)(N-2)} \sum_{h,j \in N, h \neq i, j, i \neq j} \frac{sp_{hj}(i)}{sp_{hj}}$$

where $sp_{hj}$ is the number of shortest paths between nodes $h$ and $j$, and $sp_{hj}(i)$ is the number of shortest paths between nodes $h$ and $j$ that path through $i$ [33]. Removing nodes with high $BC$ significantly impacts network performance, which is essential for efficient communication.

The networks obtained in this study are directed and weighted. Each network can potentially have at most $(N^2 - N)/2$ edges. The network sparsity, a measure of the density of connections within the network, was calculated by dividing the total number of existing edges by the maximum possible number of edges in the network. This approach allows for the comparison of different networks in terms of their connectivity density.

**Network modularity**

The modularity methods are used to evaluate the quality of a network's community structure. The goal of these methods is to identify communities, or groups of nodes, that have high internal connectivity and low external connectivity to other communities [65]. In directed networks, a modularity optimization method can be applied to identify communities [66]. This approach aims to find a division of the network into communities that maximize the benefit function $Q$, called modularity, which is defined as follows:

$$Q = \frac{1}{m} \sum_{ij} \left[ A_{ij} - \frac{k_i^{in} k_j^{out}}{m} \right] \delta_{ci,cj}$$

where $A_{ij}$ is defined conventionally: it equals 1 if there is an edge from $j$ to $i$, and 0 otherwise. The terms $k_i^{in}$ and $k_j^{out}$ represent the in-degree and out-degree of the respective vertices, whereas $m$ denotes the total number of edges in the network. $\delta_{ij} - function$ is equal to 1 if $i = j$ and 0 otherwise. Additionally, $c_i$ is the label of the community to which the vertex $i$ is assigned [66].

To identify stable communities in the brain networks of principal and non-principal groups, we employed a multistep approach. First, for each subject, we computed a 29 × 29 association matrix [67,68] via the modularity optimization method described above. The element $A_{i,j}$ in this matrix represents the number of times the nodes $i$ and $j$ are assigned to the same module across 100 runs of the above algorithm. Next, we generated a null model by randomly permuting the original partitions 100 times. For each of these 100 random partitions, we reassigned nodes uniformly to the modules present in the partition. This process yields a null model matrix, where the element $A_{ij}$ is the number of times nodes $i$ and $j$ are randomly assigned to the same community. To remove the effects of randomness, we thresholded the original association matrix by setting any element $A_{ij}$ to 0 if its value was less than the maximum value observed in the random association matrix [69]. This step ensured that only the significant co-assignments were retained in the thresholded matrix. Finally, we computed the mean of the thresholded association matrices for all the subjects within each group (principal and non-principal). This group-level thresholded association matrix served as the input for a final round of modularity optimization via the following modularity method for an undirected network [70]:

$$Q = \frac{1}{2m} \sum_{ij} [A_{ij} - P_{ij}] \delta_{ci,cj}$$

where the actual number of edges falling between a particular pair of vertices $i$ and $j$ is $A_{ij}$, and $P_{ij}$ is the expected number of edges between $i$ and $j$, a definition that allows for the possibility that there may be more than one edge between a pair of vertices, which happens in certain types of networks [70].

To quantitatively evaluate the role of each module during task execution, we utilized the integration metric as described in earlier research [71,72]. This metric measures the interactions among modules within the network. The integration of a specific module is calculated as the average number of connections that each node within that module has with nodes from other modules [73,74].

### Statistical analysis

To quantify the differences between principal and non-principal effective connectivity networks in terms of network measurements, including the clustering coefficient, local efficiency, global efficiency, betweenness centrality and network sparsity, a Rank-Sum Wilcoxon test (also known as the Mann-Whitney U test) was performed. First, we computed all network measures in each network obtained from each participant's data. In the next step, we calculated the mean of each network measure among the subjects who participated in each task (principal and non-principal) separately. The error bars represent the standard error of the mean (SEM). The same statistical analysis was performed for the behavioral data.

## Results

In our experiment, two educational multimedia programs with the same audio content were designed. The participants were randomly assigned to either the principal multimedia (P group) or the non-principal multimedia (NP group). Behavioral results were measured through the participants' performance on the NASA-TLX questionnaire and recall test. Then, main graph measurements and network modularity were computed for the cortical brain networks extracted from the EEG data. Correlation analysis was also performed between the behavioral results and graph measures to provide a better understanding of the differences in the brain patterns of connectivity between the P and NP groups. In the gamma frequency band, we did not find any statistically significant differences in the graph metrics between the P and NP networks, or any correlations between the graph metrics and the behavioral results. Therefore, results related to the gamma band have not been reported.

### NASA-TLX and recall test results

To investigate the behavioral and cognitive load differences between the P group and the NP group while they watched watching educational multimedia, we employed the recall test and the NASA-TLX. Figure 3A shows that the NASA-TLX scores of the P group were significantly higher than those of the NP group ($p = 3.2e^{-8}$). Furthermore, the accuracy of the subjects in the P group was significantly better than that in the NP group in the recall test ($p = 1.2e^{-6}$).

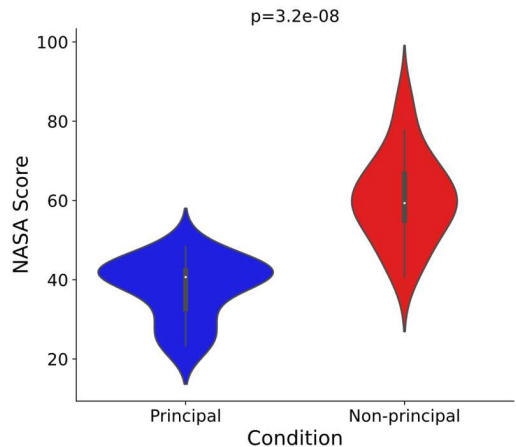
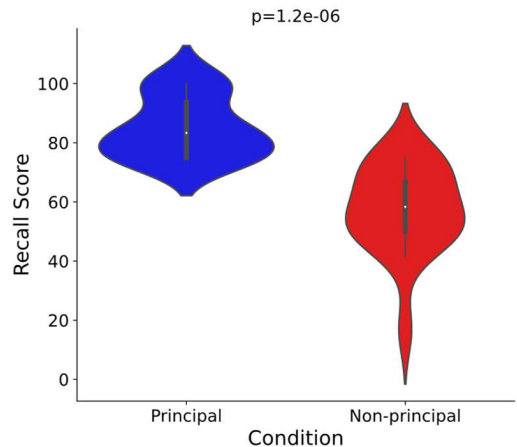

**Figure 3. Behavioral analysis: NASA-TLX scores and recall test accuracy scores**
(A) NASA-TLX scores of all the subjects after they viewed principal or non-principal multimedia. The non-principal group presented significantly higher NASA-TLX scores. (B) Accuracy scores of the recall tests for all the subjects after they watched principal or non-principal multimedia. The principal group outperformed the non-principal group. The scores were measured on a scale from 0 to 100. The p-value from the Rank-Sum Wilcoxon test is annotated in both figures.

### Graph theoretical analysis

#### Hub centers and the strength of information flow

Figure 4 shows the topographical maps of the average directionally index (DI) and total degree (TD) at each electrode during multimedia learning in both the principal and non-principal states. The connectivity patterns, as measured by TD and DI, differ between principal and non-principal states across all frequency bands.

Figure 4A displays the results of the directionality index. In the theta band, there were many connections with information flows from the posterior to anterior cortices in the NP group. A similar trend was also observed in the beta band for the NP group with fewer sender electrodes. However, for the P group, there were nonsystematic activities in the theta and beta bands. In the delta band, the flow of information was identified from the left hemisphere to the right hemisphere in the NP group. In contrast, in the P group, more receiver nodes appeared in the left hemisphere than in the right hemisphere. In the alpha band for both groups, most electrodes were sender nodes, whereas in the P group, stronger receiver nodes were located in the mFC, and in the NP group, they were found in the lateral cortices.

Concerning the total degree value, in the delta band, TD increased in the left prefrontal cortex (PFC), right parietal cortex (PC), and right temporal cortex (TC) in the NP group, whereas TD values were greater in the right and medial PC, right TC, right and medial frontal cortex (FC), and ventrolateral prefrontal cortex (VLPFC) in the P group. Compared with those in the P group, TD in the PFC, VLPFC, DLPFC and FC in the NP group increased in the theta band. In the alpha band, higher TD values were observed in the mPFC, right PFC, right VLPFC, right DLPFC, right FC, TC, and occipital cortex (OC) in the NP multimedia network, whereas the highest TD value was observed in the mFC in the P group. In the beta band, PC had higher TD values in the NP group, whereas the P group presented higher TD values in the FC.

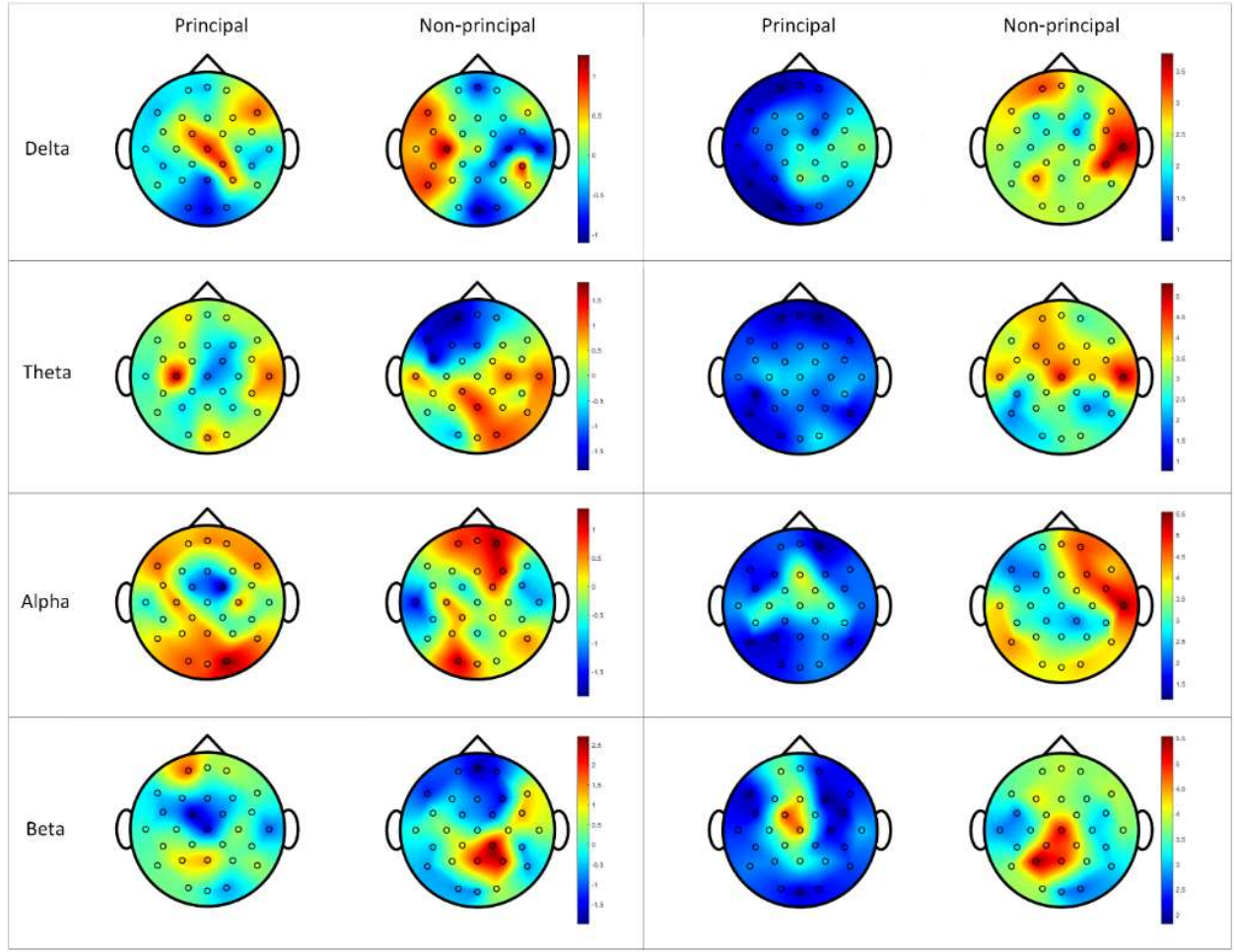

**Figure 4. Topological maps of the mean directionality index and mean total degree**
(A) Mean directionality index in principal and non-principal groups across the delta, theta, alpha, and beta frequency bands. Red indicates sender nodes, whereas blue indicates receiver nodes. (B) The mean total degree of all electrodes in the principal and non-principal brain networks through the delta, theta, alpha, and beta bands. Red indicates nodes with a high total degree, whereas blue indicates nodes with a low total degree.

Overall, in the whole brain and all frequency bands, the nodes of the NP networks have higher total degrees than the nodes of the P networks.

**Global and local information processing**
The results for local information processing or functional segregation, as measured by the clustering coefficient (CC) and local efficiency (LE), are shown in Figure 5A, B, for both the principal and non-principal groups. In the delta, alpha, and beta bands, no significant difference was observed between the mean CC values for the NP group and the P group. However, in the theta band, the mean CC value for the NP group was significantly greater than that for the P group ($p = 0.013$). Moreover, the mean local efficiency (LE) values in all bands in the P group were greater than those in the NP group. Specifically, statistically significant differences were observed in the theta and beta bands ($p = 0.009, p = 0.014$) (Figure 5B).

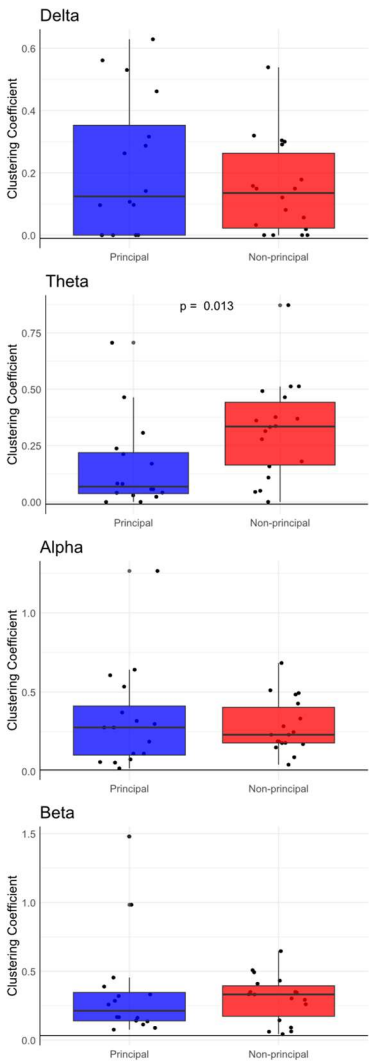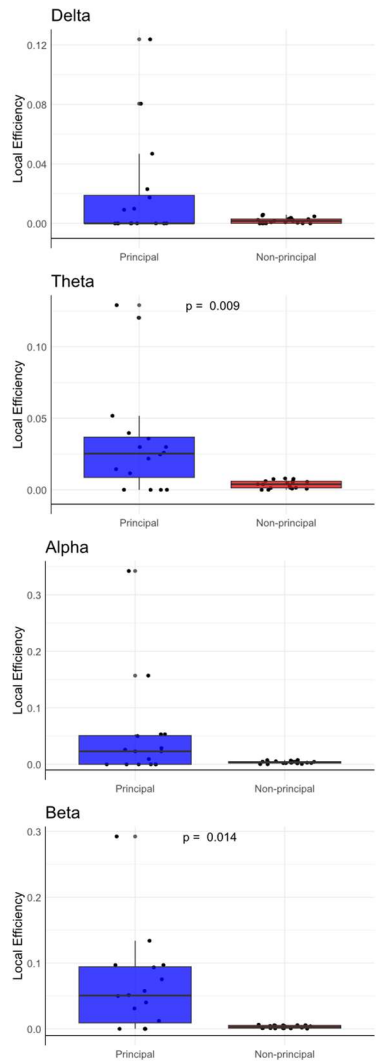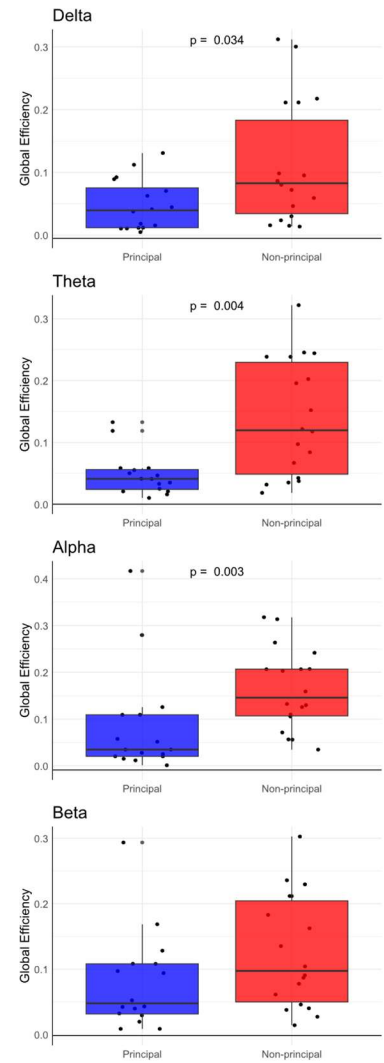

**Figure 5. Measures of functional segregation and integration**

(A, B) Functional segregation measures, clustering coefficient and local efficiency in the principal and non-principal groups across the delta, theta, alpha, and beta frequency bands. (A) The clustering coefficient of the non-principal group was significantly greater than that of the principal group in the theta band. (B) The local efficiency of the principal group was greater than that of the non-principal group in the theta and beta bands. (C) The functional integration measure, global efficiency, in the principal and non-principal groups. In the delta, theta and alpha frequency bands, the global efficiency increased significantly in the non-principal group compared with the principal group. The p-values are annotated in all figures to indicate the statistical significance of the observed differences between the two groups.

Regarding global efficiency (GE) as a representative of functional integration or global information processing, Figure 5C shows that the mean GE values in all bands in the NP group were significantly greater than those in the P group ($p = 0.034, p = 0.004, p = 0.003$), except in the beta band, where there was no significant difference.

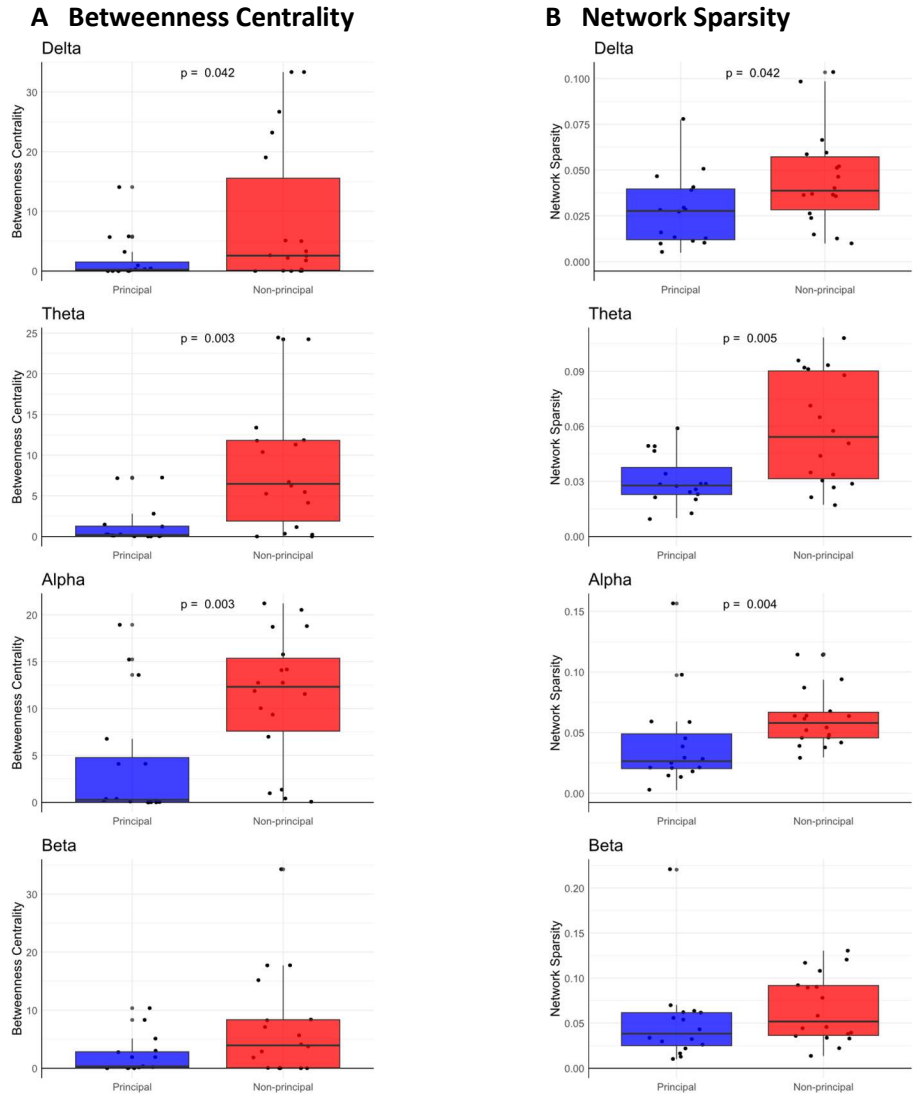

**Figure 6. Betweenness centrality and network sparsity**
(A) The mean betweenness centrality in the P and NP groups across the delta, theta, alpha, and beta frequency bands. Compared with the P group, the NP group presented significantly greater betweenness centrality values in the delta, theta, and alpha bands. (B) Network sparsity comparison. Across all frequency bands except the beta band, similar results were obtained for network sparsity, and the NP group demonstrated significantly higher network sparsity values than the principal group. The p-values are annotated in all figures.

**Measures of centrality and network sparsity**

Figure 6A shows that across all frequency bands, the NP networks have higher betweenness centrality (BC) values than the P networks do. Specifically, this difference was significant in the delta, theta, and alpha bands ($p = 0.042, p = 0.003, p = 0.003$). The results for network sparsity (NS) exhibit a similar trend. In all frequency bands, the mean value of NS in the NP group was greater than that in the P group, and statistically significant differences were observed in all frequency bands except the beta band ($p = 0.042, p = 0.005, p = 0.004$). The values of network sparsity are indicative of complex networks in the human brain [75,76], suggesting that the NP group has more complex networks than the P group does, as depicted in Figure 6B.

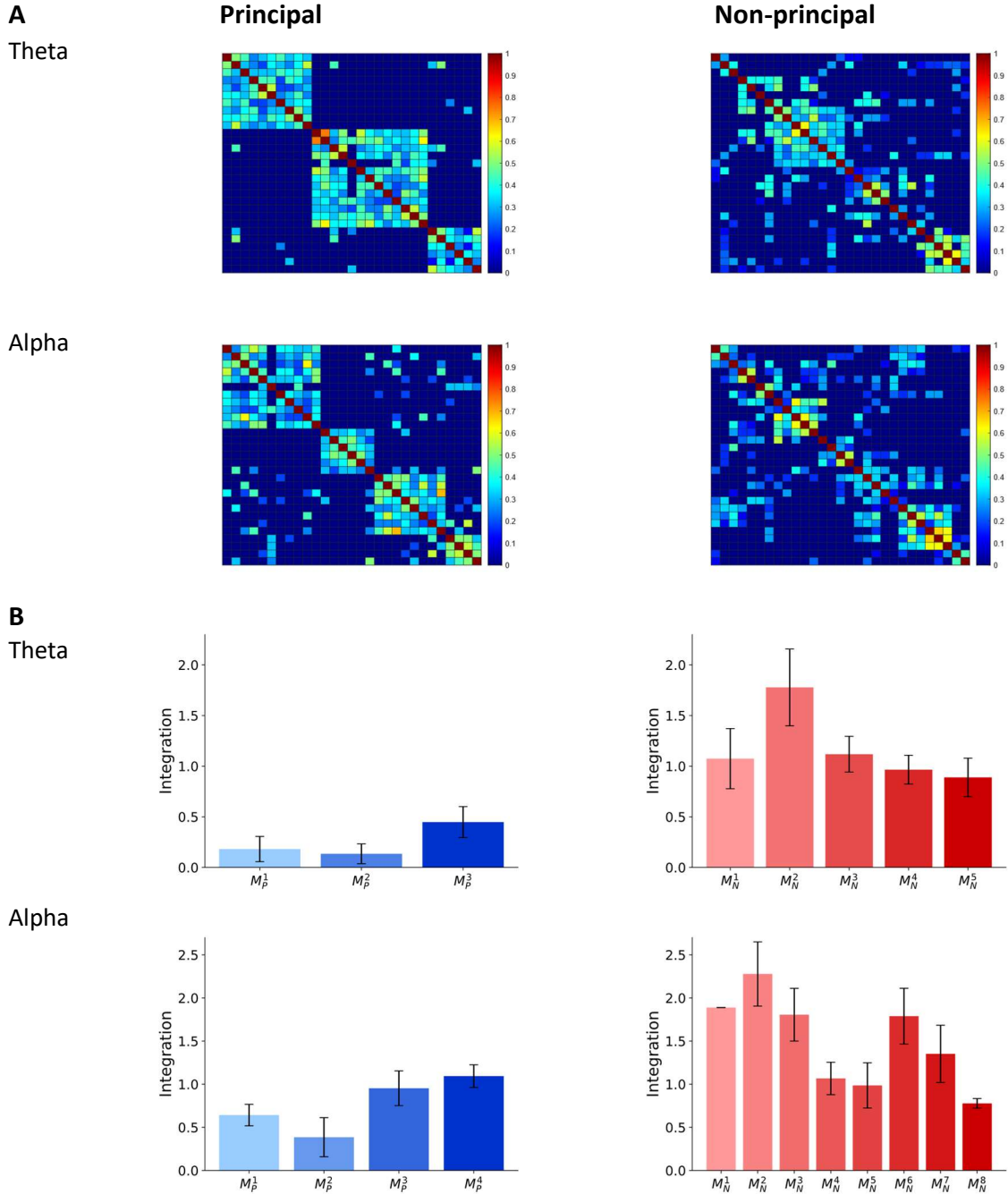

**Figure 7. Network modularity in the theta and alpha frequency bands**
(A) Modular allegiance matrices in the theta and alpha frequency bands for the principal and non-principal groups. (B) The integration values for each module in both conditions in the alpha and beta frequency bands. Compared with the principal group, the non-principal group had more modules and higher integration values in both bands.

**Network modularity**

We clustered the networks into modules or communities to further investigate the differences between the principal and non-principal groups, allowing for a distinct observation of the integration and segregation parameters. We computed network modularity for both groups in the theta and alpha frequency bands, which are particularly relevant in cognitive load studies. Specifically, we derived the final modules and interactions for each group from the group-level thresholded association matrices, as detailed in the methods section. Figure 7A shows that a qualitative visual inspection of the resulting matrices revealed distinct modular configurations between the two groups, particularly regarding intermodular activity (integration) and intramodular connectivity (segregation). To assess the average differences between the two groups, integration values (Figure 7B) were calculated for each module from the final matrices. The findings indicate that the non-principal networks exhibited high intermodular activity (high integration) and low intramodular connectivity (low segregation) in the theta and alpha frequency bands. This suggests that the non-principal group possesses a more complex and interconnected modular structure, potentially reflecting differences in cognitive processing or functional organization between the two groups.

### Correlation analysis between graph metrics and behavioral measures

We investigated the correlations between graph metrics and behavioral performance, including NASA-TLX scores and recall test accuracy, across different frequency bands, electrodes, and between the principal and non-principal groups. Significant correlations were found between the NASA-TLX scores and the graph metrics only in the theta band. However, significant correlations were observed between the recall test accuracy and the graph metrics in the alpha and beta bands (Table 1).

In the principal group, in the theta band, there was a negative correlation between local efficiency (LE) values and NASA-TLX scores at electrode C4, as well as a negative correlation between betweenness centrality (BC) values and NASA-TLX scores at electrode FC6. A positive correlation was observed between LE values at electrode CP6 and recall test accuracy in the alpha band in the principal group. Notably, all three electrodes (C4, FC6, CP6) are located in the right hemisphere (Table 1).

Regarding the non-principal group, the clustering coefficient (CC) value was negatively correlated with the NASA-TLX score in $T8$ across the theta band. As mentioned earlier, the NASA-TLX score is associated with participants' cognitive load, and this negative correlation suggests that as the value of graph measures increases, the cognitive load decreases. Recall test accuracy is significantly correlated with graph metrics in two frequency bands: alpha and beta. In the alpha band, we observed two significant negative correlations between recall test accuracy and the clustering coefficient in $P7$ and $P3$, which are closely located in the left hemisphere. Moving to the beta band, we found that the clustering coefficient was negatively correlated with recall test accuracy in $CP5$ and $T7$, as well as in the left hemisphere. Moreover, in the beta band, significant correlations were found between recall test accuracy and another graph metric, betweenness centrality, in $FP1$, $F4$, and $T8$ (Table 1).

Figure 8 shows the strongest significant correlations between graph metrics and behavioral measures in the theta and alpha frequency bands. In the alpha band, there was a positive correlation between the local efficiency of CP6 and the accuracy score of the principal group ($p = 0.01, R = 0.61$) (blue line). Conversely, there was a negative correlation between the accuracy score and the clustering coefficient of P3 in the non-principal group ($p = 0.001, R = -0.7$) (red line). In the theta band, significant negative correlations were observed at FC6 in the principal group, which links betweenness centrality to NASA-TLX scores ($p = 0.002, R = -0.71$), and at T8 in the non-principal group, which associates the clustering coefficient with NASA-TLX scores ($p = 0.01, R = -0.59$).

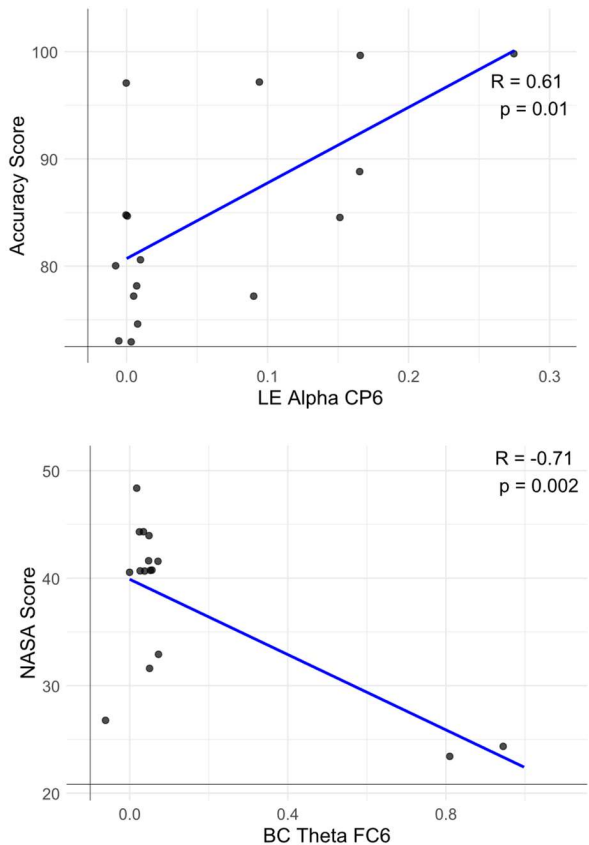 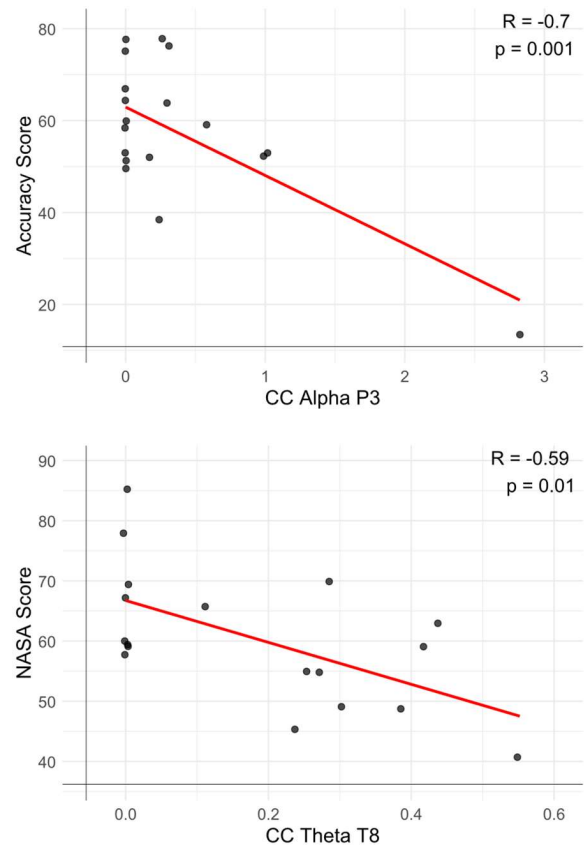

**Figure 8. Relationships between graph metrics and behavioral data**
(A) The most significant relationships between the graph metrics and the accuracy scores and NASA-TLX scores in the principal group. In the alpha band, a positive correlation was found between the local efficiency values of all the subjects in the principal group and their corresponding accuracy scores. Conversely, in the theta band, a negative correlation was observed between the betweenness centrality values of all principal subjects and their NASA-TLX scores. (B) The strongest correlations between the graph metrics and accuracy scores, along with the NASA-TLX scores, in the non-principal group. In the alpha and theta bands, significant negative correlations were found between the clustering coefficient of NP brain networks and both accuracy scores and NASA-TLX scores.

## Discussion

This study aimed to elucidate the impact of multimedia design principles on the underlying neural mechanisms supporting educational multimedia learning. In the experiment, two multimedia programs were constructed using the same auditory content but different visual designs. In one task, the visual elements adhered to Mayer's principles of multimedia learning, whereas the other intentionally violated these guidelines. The participants were randomly divided into two groups, the P group and the NP group, who watched the principal or non-principal multimedia, respectively. EEG signals were recorded from participants while watching the video and the cortical functional networks were constructed from scalp signals via the phase slope index method.

**Table 1. Correlation analysis between graph metrics and behavioral scores.**

| Condition | Frequency Band | Electrode | Graph Measurement | Behavioral Test | |
|---|---|---|---|---|---|
| **Principal** | Theta | C4 | LE | NASA-TLX | $p = 0.009, R = -0.63$ |
| | Theta | FC6 | BC | NASA-TLX | $p = 0.002, R = -0.71$ |
| | Alpha | CP6 | LE | Recall Test | $p = 0.010, R = 0.61$ |
| **Non-principal** | Theta | T8 | CC | NASA-TLX | $p = 0.010, R = -0.592$ |
| | Alpha | P7 | CC | Recall Test | $p = 0.003, R = -0.652$ |
| | Alpha | P3 | CC | Recall Test | $p = 0.001, R = -0.696$ |
| | Beta | CP5 | CC | Recall Test | $p = 0.004, R = -0.648$ |
| | Beta | T7 | CC | Recall Test | $p = 0.002, R = -0.674$ |
| | Beta | FP1 | BC | Recall Test | $p = 0.002, R = -0.676$ |
| | Beta | F4 | BC | Recall Test | $p = 0.009, R = -0.594$ |
| | Beta | P8 | BC | Recall Test | $p = 0.007, R = -0.611$ |

In this study, whole-brain network analysis allowed us to compare functional connectivity patterns evoked by the P and NP educational multimedia programs. We employed several brain network measures aligned with network modularity to explore the relationships between different brain regions, focusing on global and local connectivity measurements as well as those related to the importance of individual regions (hubs). The results indicated that principal networks were superior in local information processing, whereas non-principal networks excelled in global information processing and hub formation across various frequency bands. Modularity analysis also demonstrated that non-principal networks exhibited higher integration and lower segregation compared to principal networks, corroborating earlier findings. Moreover, significant correlations were found between participants' behavioral performance (measured by NASA-TLX scores and recall test accuracy) and network measures across different regions.

**Compensatory mechanisms in non-principal brain networks: Evidence from integration and the role of hubs**

Bassett and Sporns emphasized the importance of functional segregation, which is associated with efficient local information processing, and functional integration, which is linked to enhanced global information processing [30]. These concepts are essential for understanding brain function within the framework of network neuroscience[31–33]. Our brain network analysis indicates that, compared with P networks, NP networks exhibit increased integration, as revealed by higher global efficiency. Moreover, NP networks contain more hubs across the entire brain which facilitates extensive communication with other regions and supports functional integration, as demonstrated by higher betweenness centrality and node degree. One potential interpretation of this enhanced information processing—less observed in local processing and more prominent in global processing—is that it may represent a compensatory mechanism for efficient learning when engaging with non-principal designs.

Interestingly, this compensatory mechanism is found across the delta, theta, and alpha frequency bands in the NP networks. In these bands, the performance of NP networks is significantly enhanced by global information processing, which facilitates efficient communication throughout the entire brain. Moreover, various types of hubs, including BC and TD hubs, support this compensatory mechanism. It was also shown that NP networks contain more connectivity edges, reflected in higher network sparsity (NS) values in these three frequency bands. However, regarding local information processing, P networks outperform NP networks in the theta and beta bands when examining the ability of neighboring nodes to communicate efficiently after removing a specific node (LE). However, when considering how closely neighbors tend to cluster together (CC), a significant difference was observed only in the theta band, with higher values attributed to NP networks. This difference may explain why the neural mechanism underlying P networks leads to lower cognitive load and better performance when participants engage with multimedia designed according to Mayer's principles.

**Brain network modularity: Enhanced integration and reduced segregation in the non-principal group**

The modular architecture of human brain networks has been extensively reported via various neuroimaging techniques across different scales [77,78]. Numerous studies have demonstrated that the brain is organized into modules which play a crucial role in cognitive processes and are associated with various brain states and diseases [79,80]. This modular organization is linked to individual cognitive performance and has been shown to facilitate flexible learning and promote functional specialization [81].

In this study, we focused on the functional role of brain network modularity while participants learned from principal and non-principal multimedia. Our findings indicate that these two conditions exhibit distinct modular organizations characterized by differing integration and segregation values. Specifically, compared to principal networks, non-principal networks demonstrated higher integration and lower segregation. These results can be interpreted as an enhancement of communication between modules, driven by the increased cognitive demands of learning from non-principal multimedia. The non-organized visual information presented in NP designs may not be readily associated with familiar patterns in the brain's cognitive database, prompting participants to engage in more extensive searches for recognizable signals.

**The role of frequency bands and brain regions in the compensatory mechanism**

Previous studies have demonstrated the significant role of theta band activity in cognitive load assessment [82,83]. Our research further supports these findings through the application of network measure features. In the theta band, we observed significant negative correlations between graph metrics (local efficiency, betweenness centrality, and clustering coefficient) derived from three closely situated brain regions in the right hemisphere (C4, FC6 and T8) and NASA-TLX scores, a well-established tool for measuring subjective mental workload [11]. These negative correlations indicate that higher values of the network metrics correspond to a reduction in cognitive load. This aligns with previous findings that increased theta power during cognitive tasks is associated with better performance[82].

Similarly, research has shown that alpha-band oscillations are correlated with visual attention, selective attention and working memory [84–86]. In our study, we observed a positive correlation between the LE value of the CP6 and the performance of the subjects (accuracy in the recall tests) in the principal group. This result indicates that when local efficiency increases, accuracy also increases. This finding is consistent with research suggesting that alpha oscillations are involved in attentional processes and cognitive control [84,85]. Additionally, we found a negative correlation between the clustering coefficient values of P3 and P7 in the left hemisphere and the accuracy of subjects in the non-principal (NP) group. This suggests that individuals in the non-principal group may require higher levels of integration rather than segregation to achieve better performance. This finding aligns with previous research highlighting the significance of alpha oscillations in integrative brain function [87].

Delta band activity is also linked to various cognitive processes, including attention and memory, which are crucial for multimedia learning [88]. Delta oscillations facilitate temporal and representational integration during sentence processing, which is essential for comprehending and retaining multimedia content [89]. We found a significant increase in delta band activity within the NP group, as measured by functional integration (GE) and centrality metrics (BC and TD). These findings provide compelling evidence of compensatory mechanisms operating across different frequency bands.

**Conclusions**

The compensatory mechanism observed in NP brain networks provides valuable insights into how the brain adapts to suboptimal multimedia design. This adaptation aligns with the principles of neural plasticity and the brain's capacity to respond to cognitive challenges or limitations. Additionally, the varying effects across frequency bands highlight the intricate dynamics of brain networks during multimedia learning. Our findings indicate that NP networks exhibit compensatory mechanisms through enhanced global processing and hub formation in the delta, theta, and alpha bands, whereas principal networks demonstrate more efficient local processing, particularly in the theta and beta bands. Regarding network modularity, we also found that non-principal multimedia evoked greater integration and lower segregation in brain networks than did principal multimedia. These results underscore the importance of considering both global and local network properties in the study of brain function. Future research should delve into the temporal dynamics of these network changes and explore how individual differences in cognitive abilities may influence these compensatory mechanisms. Furthermore, integrating network analysis with other neuroimaging techniques could lead to a more comprehensive understanding of the neural underpinnings of multimedia learning.

84. Alamia, A., Terral, L., D'ambra, M. R. & VanRullen, R. Distinct roles of forward and backward alpha-band waves in spatial visual attention. *Elife* **12**, e85035 (2023).

85. Foxe, J. J. & Snyder, A. C. The role of alpha-band brain oscillations as a sensory suppression mechanism during selective attention. *Front Psychol* **2**, 154 (2011).

86. Dai, Z. *et al.* EEG cortical connectivity analysis of working memory reveals topological reorganization in theta and alpha bands. *Front Hum Neurosci* **11**, 237 (2017).

87. Başar, E. A review of alpha activity in integrative brain function: fundamental physiology, sensory coding, cognition and pathology. *International Journal of Psychophysiology* **86**, 1–24 (2012).

88. Harmony, T. The functional significance of delta oscillations in cognitive processing. *Front Integr Neurosci* **7**, 83 (2013).

89. Ding, R., Ten Oever, S. & Martin, A. E. Delta-band Activity Underlies Referential Meaning Representation during Pronoun Resolution. *J Cogn Neurosci* **36**, 1472–1492 (2024).
## Acknowledgements

We thank Kayhan Latifzadeh, Araz Farkish, and Reza Sarailoo for data acquisition, and Amir Hosein Asaadi for his advice in preparing the figures and his help in the initial data analysis.
## Author contributions

M.O.V., M.G., A.B., S.A., and R.E. conceived, developed, and refined the central idea of the present study. M.O.V analyzed the data. M.O.V., M.G. and R.E. discussed the results. M.O.V. and M.G. visualize the results. M.G. wrote the manuscript, and M.O.V. and R.E. provided critical revisions. All authors approved the final version of the manuscript.

## Data availability statement

The datasets generated during the current study are available from the corresponding author on reasonable request; please send a request to ebrahimpour@sharif.edu.

## Additional Information

**Competing financial interests:** The authors declare no competing financial interests.